\documentclass[a4paper,nofootinbib, aps]{revtex4}

\usepackage{color}
\usepackage{amsmath,amsfonts,amssymb,amsthm}
\usepackage{graphicx}
\usepackage{fullpage}
\usepackage{subcaption}
\usepackage{xfrac}

\usepackage{tikz}
\usetikzlibrary{arrows,decorations.pathmorphing,backgrounds,positioning,fit}
\usepackage{epstopdf} % to include .eps graphics files with pdfLaTeX
\usepackage{bm}  % Define \bm{} to use bold math fonts

\usepackage[pdfpagelabels,pdftex,bookmarks,breaklinks]{hyperref}
\definecolor{darkblue}{RGB}{0,0,127} % choose colors
\definecolor{darkgreen}{RGB}{0,150,0}
\hypersetup{colorlinks, linkcolor=darkblue, citecolor=darkgreen, filecolor=red, urlcolor=blue}

\usepackage{layout}

\def\Z{\mathbb{Z}}
\def\R{\mathbb{R}}
\def\C{\mathbb{C}}

\newcommand{\Eref}[1]{Eq.~(\ref{#1})}
\newcommand{\Sref}[1]{Sec.~\ref{#1}}
\newcommand{\Fref}[1]{Fig.~\ref{#1}}

\def\th{^{\rm th}}

\newcommand{\ket}[1]{|{#1}\rangle}

\newcommand{\ketbra}[2]{|{#1}\rangle\!\langle{#2}|}

\newcommand{\proj}[1]{\ketbra{#1}{#1}}

\begin{document}

\title{Generalized Cluster States Based on Finite Groups}

\author{Courtney G.\ Brell}
\email{courtney.brell@gmail.com}
\affiliation{Centre for Engineered Quantum Systems, School of Physics, The University of Sydney, Sydney, Australia}
\affiliation{Institut f\"ur Theoretische Physik, Leibniz Universit\"at Hannover, Appelstra\ss e 2, 30167 Hannover, Germany}

%\date{\today}

\begin{abstract}
We define generalized cluster states based on finite group algebras in analogy to the generalization of the toric code to the Kitaev quantum double models. We do this by showing a general correspondence between systems with CSS structure and finite group algebras, and applying this to the cluster states to derive their generalization. We then investigate properties of these states including their PEPS representations, global symmetries, and relationship to the Kitaev quantum double models. We also discuss possible applications of these states.
\end{abstract}

\maketitle

%------------------------------------------------------------------------------------------------------------%

\def\L{\leftarrow}
\def\R{\rightarrow}

\section{Introduction}

Cluster states (or graph states)~\cite{Raussendorf2001, Raussendorf2003} are the prototypical resource for measurement-based quantum computation (MBQC)~\cite{Briegel2009}. This is broad category of strategies for implementing fault-tolerant quantum information processing equivalent but in contrast to the quantum circuit model, adiabatic quantum computation, and topological quantum computation. MBQC proceeds by taking a suitable entangled many-particle resource state and performing computation by sequential single-particle measurements that may be chosen adaptively based on previous measurement results. 

The cluster states have a particularly simple structure that allows for straightforward analysis of their properties. They also have many desirable features for a resource state: for example, they are the output of a finite-depth quantum circuit, and also the frustration-free ground states of a (gapped) commuting local Hamiltonian.

Apart from their usefulness for standard MBQC, the cluster states are also related to topologically ordered systems such as the toric code~\cite{Kitaev2003,Dennis2002, Raussendorf2005,Han2007,Brown2011} and the color codes~\cite{Bombin2006,Bombin2008}. In fact, this relationship can be leveraged to define a protocol for MBQC that exploits the natural fault-tolerance properties of topological quantum computation schemes~\cite{Raussendorf2007}. The cluster states can also be used to study general stabilizer states, as all stabilizer states are equivalent to cluster states under local Clifford operations~\cite{VandenNest2004}. Additionally, the cluster states have been used in studies of such diverse topics as the origin of quantum computational power~\cite{Anders2009} and contextuality~\cite{Raussendorf2013}.

Since the cluster states are such an important theoretical tool for studying these topics, it is of interest to ask if their properties can be generalized in any interesting ways. In this paper we take steps towards answering this question, in particular motivated by the relation between cluster states and topologically ordered systems. In so doing, we provide a general framework for similar generalization programmes.

The toric code is the simplest member of the family of topologically ordered Kitaev quantum double models~\cite{Kitaev2003}. These models are defined by a finite group $G$ (of which the toric code corresponds to $\Z_2$). These models are of significant interest for condensed matter physics, where they are an important testbed for the phenomenology of topological order, as well as quantum information, where they can be used to implement topological quantum computation, for example through braiding of quasiparticles~\cite{Nayak2008} or code deformation~\cite{Bombin2009}, or can be used as quantum memories~\cite{Dennis2002}. Though extensions of the Kitaev quantum double models have also been proposed based on a Hopf algebra $H$~\cite{Buerschaper2013}, or twisted quantum doubles labelled by finite group $G$ and 3-cocycle $\omega$~\cite{Hu2013}, we will restrict our consideration to the finite group case here. Most of the important phenomenology of these generalizations can be captured by finite groups, and in particular the qualitative distinction between Abelian and non-Abelian topological phases is manifested in Kitaev quantum double models for Abelian and non-Abelian groups respectively. Abelian phases such as the toric code cannot be used for quantum computation by braiding of quasi-particles, and are not known to be able to implement universal topological quantum computation via code-deformation (though universal quantum computation can still be achieved by using non-topological operations such as magic state distillation~\cite{Bravyi2005,Cesare2014}).

For these reasons, we define a family of generalized cluster states based on arbitrary finite groups $G$, where the standard qubit cluster state corresponds to the simplest group $\Z_2$. The previously known higher-qudit cluster states~\cite{Zhou2003, Hall2007} correspond to the cyclic groups $\Z_d$. In order to make this generalization process as clear as possible, we first make a general connection between general qubit CSS states and the group $\Z_2$. This allows an intuitive generalization of any such system, which may be more generally applicable.

We explore the properties of the generalized cluster states defined in this way, such as their global symmetries and PEPS representations, in analogy to the qubit case~\cite{Son2011,Verstraete2004}. We also show how the generalized cluster states retain a relation to the corresponding quantum double models, and discuss possible applications of the generalized cluster states making use of this relation.

The paper is organized as follows. In \Sref{s:css} we outline the group structure of CSS states and the general method of generalization we will follow. This is then used in \Sref{s:cs} to define the generalized cluster states for arbitrary finite groups $G$. \Sref{s:properties} is devoted to exploring some properties of these states. Finally, in \Sref{s:discuss} we briefly consider applications for generalized cluster states including preparation of the Kitaev quantum double states and a generalization of the topological cluster state computation scheme, before discussing possible extensions of the kind of generalization scheme proposed here and providing some concluding remarks.

%------------------------------------------------------------------------------------------------------------

\section{From CSS structure to finite groups}\label{s:css}

	In this paper, we will generalize the familiar qubit cluster state to states based on finite group algebras. In order to do this, we will make use of a general prescription to translate from a system with CSS structure~\cite{Calderbank1996, Steane1996} to one based on the group $\Z_2$.

	Many interesting models defined on spin-$\sfrac{1}{2}$ systems have a CSS structure. While the term CSS has a specific technical meaning as a class of stabilizer codes, we will use it in its more colloquial sense to mean any system involving interactions that consist either of products of Pauli $X$ operators or products of Pauli $Z$ operators.

	Systems with a CSS structure have a natural interpretation in terms of the group algebra of $\Z_2$. Recall that $\Z_2$ has two elements (labelled $1$ and the identity $0$), group multiplication is addition modulo 2 ($\oplus$), and there are two irreducible representations. The trivial and alternating irreps we label $+$ and $-$ respectively, with $\Gamma_+(0)=\Gamma_+(1)=\Gamma_-(0)=1$ and $\Gamma_-(1)=-1$.
	
	We can associate qubit states and operators with objects related to the group $\Z_2$ by considering the computational basis states of our qubit to be labelled by group elements. That is, we take $\ket{0}$ and $\ket{1}$ to be associated with the respective group elements $0$ and $1$. Following this, the Pauli $X$ operator can be seen to act as group multiplication by the $1$ element. For this reason we denote it by $X_1\equiv X$. We can also see that $X_0\equiv X^0=I$ acts as group multiplication by the $0$ element. Thus we consider $X_g$ to act as group multiplication by an element $g\in\Z_2$. We can also interpret the CNOT gate in this context as a controlled group multiplication operation. This can easily be seen by $\mathrm{CNOT}\ket{g}\ket{h}=\ket{g}\ket{g\oplus h}$.
	
	The conjugate basis states $\ket{+}$ and $\ket{-}$ can be associated with the irreducible representations of $\Z_2$ by noticing that $\ket{\pm}\propto\Gamma_{\pm}(0)\ket{0}+\Gamma_{\pm}(1)\ket{1}$ (throughout this paper we will consider only unitary irreducible representations over $\C$ unless otherwise specified). We can similarly consider the powers of the Pauli $Z$ operator to be associated with the representations of $\Z_2$ as $Z_+\equiv Z^1$ and $Z_-\equiv Z^0$. Notice then that $Z_{\pm}\ket{g}=\Gamma_{\pm}(g)\ket{g}$.
	
	The group $\Z_2$ has much structure that is absent for general groups. In particular, there is a natural isomorphism between group elements and irreps, taking $0$ to $+$ and $1$ to $-$. In this way we can interpret the Hadamard gate as implementing this isomorphism. Although the CPHASE gate has no fundamental interpretation in terms of objects from the group, it can be brought into this framework by noticing that CPHASE can be constructed from a circuit of Hadamard and CNOT gates.
	
	This entire structure is summarized in the first two columns of Table \ref{t:css-qd}. It also gives us an avenue to generalize many of these concepts to arbitrary groups. In interpreting the structure of the qubit in terms of $\Z_2$, the most significant property of this group that does not still hold in the general case is the existence of a natural isomorphism from group elements to irreps. The fact that this is not available for a general group $G$ is the reason that the CSS structure is important when generalizing to arbitrary finite groups, as we will show.
	
	Now that the properties of the qubit have been related to properties of $\Z_2$, the generalization to an arbitrary finite group $G$ is relatively straightforward. 2-dimensional qubits are replaced by $|G|$-dimensional qudits. The analogue of computational basis states for these qudits are labelled by group elements $g\in G$ (we call this the group element basis). It is worthwhile noting that there is a preferred state corresponding to the identity group element $e$. Thus the $\ket{0}$ state of the qubit corresponds in general to the $\ket{e}$ state of the group element basis.
	
	 The Pauli $X$ operator of the qubit generalizes to left and right group multiplication operators $X_g^{\L}$ and $X_g^{\R}$ acting as
	\begin{align}
		X_g^{\L}\ket{h}&=\ket{gh}\;\;\mbox{and}\;\;X_g^{\R}\ket{h}=\ket{hg^{-1}}.
	\end{align}
	Similarly, the CNOT gate generalizes straightforwardly to a controlled left or right group multiplication gate. 	
	
	The conjugate basis of the qubit was interpreted as representation states, and in the general case we label these states by matrix elements of an unitary irreducible representation (or irrep) of $G$, and define
	\begin{align}
		\ket{\Gamma^{ij}}& = \sqrt\frac{d_{\Gamma}}{|G|}\sum_{g\in G}[\Gamma(g)]_{ij}\ket{g},
	\end{align}
	where $[\Gamma(g)]_{ij}$ is given by the group element $g$ at a matrix element $(i,j)$ of a representation $\Gamma$ and $d_{\Gamma}$ is the dimension of $\Gamma$. We call this basis the representation basis as compared to the group element basis. Noting that $\sum_{\Gamma}d_{\Gamma}^2=|G|$, orthonormality of the representation basis follows directly from the grand orthogonality theorem of group representations:
	\begin{align}
		\sum_{g\in G}[\Gamma_{\lambda}(g)]^*_{ij}[\Gamma_{\sigma}(g)]_{i'j'}& = \delta_{\lambda\sigma}\delta_{ii'}\delta_{jj'}\frac{|G|}{d_{\lambda}}.
	\end{align}

	The transformation from the representation basis to the group element basis can also be found as
	\begin{align}
		\ket{g}&= \sum_{\Gamma^{ij}}\sqrt\frac{d_{\Gamma}}{|G|}[\Gamma(g)]_{ij}\ket{\Gamma^{ij}}.
	\end{align}
	
	As in the group element basis, there is a preferred state in the representation basis. This corresponds to the trivial irrep $I$. The trivial irrep state is given by an equal superposition over all group element states, and reduces to the $\ket{+}$ state in the case that $G=\Z_2$.

	The generalizations of the Pauli $Z$ operators are labelled by matrix elements of irreps of $G$. As in the $\Z_2$ case, they act on group element basis states by accumulating amplitudes corresponding to the relevant matrix element of the group element in the given representation. That is,
	\begin{align}
		Z_{\Gamma^{ij}}\ket{h}=[\Gamma(h)]_{ij}\ket{h}.
	\end{align}
	These generalizations are summarized in Table \ref{t:css-qd}.
	
	The Hadamard gate and CPHASE gate of the qubit have no natural generalization to an arbitrary group in this way. Similarly, interactions that do not have CSS structure cannot be generalized for the following reason. If we consider an operator constructed from mixed $X$ and $Z$ operators on a set of qubits, to generalize this operator to the group $G$ we must associate a group element to the $X$ operator and a representation to the $Z$ operator. There is no natural way to choose a representation corresponding to each group element, and so we cannot consistently generalize a mixed operator of this type. This is a direct consequence of the fact that the Hadamard operator has no analogue for arbitrary groups. Given that we can only consider systems with CSS structure in this framework, it is clear that the Hadamard and CPHASE gates cannot be generalized, as they take CSS systems to non-CSS systems.
	
	One way to interpret Table \ref{t:css-qd} is as a prescription to design quantum systems with algebraic properties inherited from an arbitrary finite group. In general, the structure of these algebraic properties will be closely related to the quantum double of the group under consideration, as is the case for example in the Kitaev quantum double models, which can be interpreted in this framework. One could also consider generalizing this correspondence from groups to more general objects such as Hopf algebras (as has been done for the Kitaev quantum double models). This will briefly be discussed in Section \ref{s:extension}, as well as speculation on further possible generalizations.
	
	\begin{table}[t]
	\centering
	\begin{tabular}{c|c||c|c}
		 $\mathbb{Z}_2$&  Qubit& Qudit&  $G$ \\
	  	\hline\hline
	  	Elements $\{0,1\}$&$\ket{0},\ket{1}$&$\ket{g}$~~($d=|G|$)&Elements $g\in G$\\\hline
	  	Multiplication $\oplus g$&$X_g\ket{h} = \ket{g\oplus h}$ &\begin{tabular}{c}$X_g^{\L}\ket{h} =\ket{gh}$\\$X_g^{\R}\ket{h} = \ket{hg^{-1}}$\end{tabular}&Multiplication\\\hline
	  	Irreps $\{\Gamma_+,\Gamma_-\}$&$\ket{\pm}=\frac{1}{\sqrt{2}}\left(\Gamma_{\pm}(0)\ket{0}+\Gamma_{\pm}(1)\ket{1}\right)$&$\ket{\Gamma^{ij}} = \sqrt\frac{d_{\Gamma}}{|G|}\sum_g [\Gamma(g)]_{ij}\ket{g}$& Irreps $\Gamma\in \mathrm{Rep}(G)$\\\hline
	  	Irrep Action $\Gamma(g)$&$Z_{\Gamma}\ket{h}=\Gamma(h)\ket{h}$&$Z_{\Gamma^{ij}}\ket{h}=[\Gamma(h)]_{ij}\ket{h}$& Irrep Action $[\Gamma(g)]_{ij}$\\\hline
	  	$\mathbb{Z}_2\cong \mathrm{Rep}(\mathbb{Z}_2)$&Hadamard gate&-&-\\\hline
	  	Controlled Not& $\mathrm{CNOT}\ket{g}\ket{h}=\ket{g}\ket{g\oplus h}$& \begin{tabular}{c}$\mathrm{CMULT}^{\L}\ket{g}\ket{h}=\ket{g}\ket{gh}$\\$\mathrm{CMULT}^{\R}\ket{g}\ket{h}=\ket{g}\ket{hg^{-1}}$\end{tabular} & Controlled Multiplication
	\end{tabular}
	\caption{\label{t:css-qd}Summary of the correspondence between group algebras and states or operators on quantum systems.}
\end{table}

	Of course, the machinery introduced here simplifies significantly if $G$ is chosen to be a cyclic group $\Z_d$. In this case, the representations are all 1-dimensional $d\th$ roots of unity. The basis change between the group element and representation bases for these groups is simply the discrete Fourier transform. Significantly, there also exists a natural isomorphism from the space of irreps of $\Z_d$ to the space of elements. This allows us to generalize the Hadamard gate and removes the necessity of considering only systems with CSS structure. The cluster states corresponding to the cyclic groups have been previously defined and studied~\cite{Zhou2003,Hall2007}.
	
%------------------------------------------------------------------------------------------------------------%

\section{Generalized cluster states}\label{s:cs}

	We can use the general machinery established in the previous section to generalize the cluster state. This will give a family of states labelled by decorated graphs and a group $G$. As a prelude to the introduction of these general states, let us first review the definition of the qubit cluster state. Since the standard definition does not have CSS structure, we will need to consider a slightly modified definition of the cluster state that is more amenable to generalization.

	\subsection{Qubit cluster states}\label{s:qubitcs}
	
	A qubit cluster state~\cite{Raussendorf2001, Raussendorf2003} is uniquely specified by an underlying graph $\Lambda$. One convenient way to define these states is as the output of a certain constant depth quantum circuit. Equivalently, they can be described as the common $+1$ eigenstate of a set of commuting stabilizer operators. We will describe the qubit cluster state in both of these languages, as these two descriptions showcase different desirable features of the cluster state. In the first case, the cluster state can be prepared in constant time by an appropriate parallel quantum circuit, while in the latter case the cluster state can be seen as the unique gapped ground state of a local Hamiltonian, namely that formed by the (negative) sum of the relevant stabilizer operators.
	
	The cluster state is given by
	\begin{align}\label{e:qubitcirc}
		\ket{\mathcal{C}_{\Lambda}} = \prod_{<m,n>}\mathrm{CPHASE}(m,n)\bigotimes_{v\in \Lambda}\ket{+}_v
	\end{align}
	with $<m,n>$ running over all edges and $v$ running over all vertices of $\Lambda$. The CPHASE gate is given by $\mathrm{CPHASE}(m,n)\ket{a}_m\ket{b}_n = (-1)^{ab}\ket{a}_m\ket{b}_n$ for $a,b\in \Z_2$. As a circuit, we place qubits in the $\ket{+}$ state at each site of $\Lambda$ and then perform CPHASE gates between the qubits connected by an edge. Since the CPHASE gates commute, this is always a constant-depth circuit (assuming bounded degree of $\Lambda$).
	
	It is straightforward to derive the stabilizers of this state by considering the circuit of \Eref{e:qubitcirc} in the Heisenberg representation. The qubit cluster state is thus the common $+1$ eigenstate of the stabilizer operators
	\begin{align}\label{e:qubitstabs}
		S(v) = X(v)\prod_{w\sim v}Z(w)
	\end{align}
	for every site $v$, where $w\sim v$ runs over neighbours of $v$. Clearly, this state does not have the CSS structure discussed in \Sref{s:css}. This can be seen in two ways: firstly the stabilizers (\ref{e:qubitstabs}) involve both $X$ and $Z$ operators in the same stabilizer, and secondly the circuit (\ref{e:qubitcirc}) involved CPHASE operators. The toolkit introduced in \Sref{s:css} cannot be used to generalize states of this form.
	
	We can, however modify the definition of the cluster state to endow it with CSS structure. In order to do this, we must restrict to bipartite graphs $\Lambda$. On these graphs, we can partition the sites of $\Lambda$ into an odd set $\Lambda_o$ and an even set $\Lambda_e$ such that all edges of $\Lambda$ involve one vertex from $\Lambda_o$ and one vertex from $\Lambda_e$. Given this structure, we can define the CSS cluster state as
	\begin{align}
	\ket{\mathcal{C}_{\Lambda}^{CSS}}\equiv \prod_{v\in\Lambda_e} H(v) \ket{\mathcal{C}_{\Lambda}},
	\end{align}
	where $H$ is the Hadamard operator. This allows us to rewrite the circuit constructing the CSS cluster state as
	\begin{align}\label{e:csscirc}
	\ket{\mathcal{C}_{\Lambda}^{CSS}} = \prod_{\substack{<m,n>\\ m\in \Lambda_o, n\in \Lambda_e}}\mathrm{CNOT}(m,n)\bigotimes_{w\in \Lambda_o}\ket{+}_w\bigotimes_{v\in \Lambda_e}\ket{0}_v
	\end{align}
	where here the CNOT gates act with the odd qubit $m$ as control and even qubit $n$ as target. We will generally use the notation that a controlled gate takes two site labels as arguments and acts on the first as the control. The bipartite property of $\Lambda$ guarantees that each CNOT gate acts on one odd and one even qubit. Even though CNOT gates do not commute in general, two CNOT gates with common targets or common controls will commute. This gives us the fact that the circuit specified by (\ref{e:csscirc}) is again always constant-depth as expected.
	
	The stabilizers of the CSS cluster state are given by
	\begin{align}
		S^e(v) &= Z(v)\prod_{w\sim v}Z(w)\mbox{, and}\\
		S^o(w) &= X(w)\prod_{v\sim w}X(v)
	\end{align}
	for $v$ and $w$ even and odd sites respectively. This demonstrates the claimed CSS structure of these states. This can also be seen by noting that the circuit (\ref{e:csscirc}) consists only of operators that preserve CSS structure (i.e.~CNOT gates). Although it can only be defined on bipartite graphs, when it exists the CSS cluster state is locally equivalent to the corresponding standard cluster state and so has all the same fundamental properties, notably including the ability to perform universal measurement-based quantum computation.

%------------------------------------------------------------------------------------------------------------%
	\subsection{Finite group cluster states}\label{s:groupcs}
		
%------------------------------------------------------------------------------------------------------------%

	Given the toolkit of \Sref{s:css} and the definition of the CSS cluster state given above, we will now describe the generalization of the cluster state to arbitrary finite groups $G$. These cluster states will inherit many of the algebraic properties of the group that defines them. 
	
\begin{figure}
\centering
\includegraphics{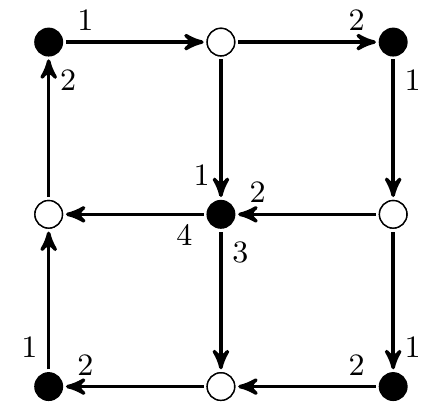}
\caption{A directed bipartite graph augmented with an ordering of edges incident to each even vertex. Even vertices are shown as solid circles, while odd vertices are represented by open circles. This data is sufficient to specify a generalized cluster state.}
\label{f:egcluster}
\end{figure}

	As compared to the qubit cluster states where an undirected graph completely specifies the state, for an arbitrary group $G$ this is insufficient to uniquely determine a generalized cluster state. This is similar to the generalization of the toric code to the quantum double models, where the former is completely specified by an undirected graph, while the latter requires directed edges to specify the model. In particular, to describe a generalized cluster state we require a directed bipartite graph $\Lambda$, together with an ordering $\#_v(w)$ of the neighbours $w$ of each even vertex $v$. An example of such a structure is shown in \Fref{f:egcluster}. The roles that are played by the additional direction and ordering parameters will become clear during the construction of the generalized cluster states.
	
	Given such a graph $\Lambda$ and ordering $\#$, together with a finite group $G$, the generalized cluster state is a state on a system of $|G|$-dimensional qudits at each site of $\Lambda$. It is most convenient to generalize the qubit cluster state by considering the circuit that defines it according to \Eref{e:csscirc}. The elements in this circuit are CNOT gates, $\ket{+}$ states and $\ket{0}$ states, each of which has a natural analogue in Table \ref{t:css-qd}.
	
\subsubsection{Circuit representation}

	The qubit $\ket{+}$ state is interpreted as a special case of the trivial irrep state $\ket{I}=\frac{1}{|G|}\sum_g\ket{g}$. Similarly the $\ket{0}$ state corresponds to the identity group element state $\ket{e}$ in general. Finally, the last ingredient in the CSS cluster state preparation circuit is the CNOT gate, which generalizes to either the left or right controlled multiplication gates $\mathrm{CMULT}^{\leftrightarrows}$.
	
	Loosely speaking then, we define the generalized cluster state as
	\begin{align}\label{e:almostdefcs}
		\ket{\mathcal{C}_{\Lambda, \#, G}^{CSS}}\sim \prod_{\substack{<m,n> \\m\in \Lambda_o, n\in \Lambda_e}}\mathrm{CMULT}(m,n)\bigotimes_{w\in \Lambda_o}\ket{I}_w\bigotimes_{v\in \Lambda_e}\ket{e}_v.
	\end{align}	
	
	There are two ways in which (\ref{e:almostdefcs}) does not yet constitute a full specification of the state $\ket{\mathcal{C}_{\Lambda, \#, G}^{CSS}}$. The first is that there is an ambiguity regarding whether CMULT gates should act as left or right multiplication. The second is that we have not specified an order in which to apply the CMULT gates, which in general do not commute on their targets. Resolving these ambiguities requires the additional structure we have specified in the $\#$ orderings and in the directions of the edges of $\Lambda$. Note that since the CMULT gates commute on their controls, we need not be concerned about the relative order of gate application on odd qudits, and so need not specify a global order of CMULT application. It suffices to specify the ordering separately at each even vertex. 
	
	We can complete the specification of \Eref{e:almostdefcs} by firstly choosing the ordering of the CMULT gates at a given even site $v$ according to $\#_v$. Secondly, we choose the $\mathrm{CMULT}(m,n)$ gate to act as left multiplication if the edge $(m,n)$ runs from $n$ to $m$, and as right multiplication otherwise. This convention can be remembered by placing the control (odd) qudit on the left of the target (even) qudit. The multiplication sense is then given directly by the direction of the edge connecting the qudits. We denote these conventions in the following way:
	\begin{align}\label{e:defcsgeneral}
		\ket{\mathcal{C}_{\Lambda, \#, G}^{CSS}}\equiv \sideset{}{^\#}\prod_{\substack{<m,n> \\m\in \Lambda_o, n\in \Lambda_e}}\mathrm{CMULT}^{\leftrightarrows}({m,n})\bigotimes_{w\in \Lambda_o}\ket{I}_w\bigotimes_{v\in \Lambda_e}\ket{e}_v
	\end{align}	
	where $\mathrm{CMULT}^{\leftrightarrows}(m,n)$ acts as $\mathrm{CMULT}^{\rightarrow}$ ($\mathrm{CMULT}^{\leftarrow}$) for an edge directed from $m$ to $n$ ($n$ to $m$). This is now a complete specification of the generalized cluster state $\ket{\mathcal{C}_{\Lambda, \#, G}^{CSS}}$. Note that since the CMULT operators commute on their controls, the circuit (\ref{e:defcsgeneral}) is always constant-depth as in the qubit case (again assuming $\Lambda$ has bounded degree).
	
	Strictly speaking, the orderings $\#$ contain more information than is necessary to specify the state $\ket{\mathcal{C}_{\Lambda, \#, G}^{CSS}}$. Since left multiplication commutes with right multiplication, we could rearrange the order of CMULT application without affecting the state $\ket{\mathcal{C}_{\Lambda, \#, G}^{CSS}}$ as long as we do not change the relative order of CMULT gates acting on edges directed outwards and the relative order of gates acting on edges directed inwards from a given even vertex. For notational simplicity we will continue to use the redundant description $\#$ of the ordering at each even vertex.
	
	The circuit (\ref{e:defcsgeneral}) is a perfectly adequate definition of the generalized cluster states. However, the stabilizer formalism has proved to be an extremely powerful tool in the analysis of cluster states and similar qubit systems. For this reason we will compute the stabilizer representation of the generalized cluster states.

	\subsubsection{Stabilizer representation}
		
		The term ``stabilizer'' has slightly different meanings in different contexts. Often it is used to mean a subgroup of the $n$-qubit Pauli group that does not contain $-1$. The relevant stabilizer states are then the common $+1$ eigenstates of the stabilizer group. This is the sense in which the qubit cluster state and the CSS cluster state we discussed in \Sref{s:qubitcs} are stabilizer states of their relative stabilizer groups.
		
		More generally, stabilizer states can be thought of as the $+1$ eigenstates of an arbitrary set of operators. Varying degrees of structure can be introduced to make the set more amenable to study, for example requiring that the stabilizer operators are monomial matrices as in the monomial stabilizer formalism~\cite{VandenNest2011}. As with the stabilizers of the Kitaev quantum double models, the generalized cluster states can be cast in the monomial stabilizer formalism. We may abuse the language slightly by referring to a set of stabilizers in the generalized sense as a group, even if we do not present them in a form that has the algebraic structure of a group.
		
		One of the most attractive features of the Pauli stabilizer formalism is the ability to efficiently simulate operations that map stabilizer states to stabilizer states (through the Gottesman-Knill theorem). This ability carries directly over to stabilizer formalisms based on Abelian group algebras~\cite{VandenNest2013efficient,Bermejo2014classical,Bermejo2014normalizer}. Such strong simulability properties seem to require more structure than is available for general stabilizer states. However, some notions of classical simulability remain in certain cases. For example, for sufficiently structured states such as the Kitaev quantum double models, the monomial stabilizer formalism allows the efficient evaluation of expectation values for local observables~\cite{VandenNest2011}. We expect analogous simulability results to apply to our models as they are based on the same algebraic structure.

		Apart from the motivation of classical simulation, the existence of a local stabilizer description guarantees the existence of a local, commuting, gapped parent Hamiltonian for these states. It can also be a useful way of illustrating phenomena that would be more difficult to describe in the Schrodinger representation, for example as is the case in the Kitaev quantum double models where stabilizer operators correspond directly to quantities of interest such as charges or Wilson loops.
		
		In order to explicitly determine a set of stabilizer operators for the generalized cluster states, we examine the action of the circuit implied by \Eref{e:defcsgeneral} in the Heisenberg representation. This circuit begins with a product state of $\ket{I}$ on odd vertices and $\ket{e}$ on even vertices. 
		
		Now, notice that the state $\ket{I}=\frac{1}{\sqrt{|G|}}\sum_g\ket{g}$ is stabilized by all group multiplication operators. That is,
		\begin{align}
			X_g^{\leftrightarrows}\ket{I}&=\ket{I}
		\end{align}
		for all $g$, and either multiplication sense $\leftrightarrows$. Similarly, we can define a set of stabilizers for $\ket{e}$, noting that
		\begin{align}
			Z_{\Gamma^{ii}}\ket{e}&=\ket{e}
		\end{align}
		for all $\Gamma,i$.
		
		Although this is a perfectly legitimate stabilizer description of the state $\ket{e}$, for the purposes of calculating the stabilizers of the state $\ket{\mathcal{C}_{\Lambda, \#, G}^{CSS}}$ it will be more convenient to use an alternative representation. Defining the projectors to group element basis states $T_g\equiv\proj{g}$, it is clear that $\ket{e}$ is stabilized by $T_e$. A more roundabout way to see this is to write the $T_g$ operators in terms of the $Z_{\Gamma^{ij}}$ operators as
		\begin{align}
			T_g&=\frac{1}{|G|}\sum_{\Gamma}d_{\Gamma}\sum_{i,j}[\Gamma(g)]_{ij}^*Z_{\Gamma^{ij}}
		\end{align}
		and find
		\begin{align}
			T_e\ket{e}&=\frac{1}{|G|}\sum_{\Gamma}d_{\Gamma}\sum_{i,j}[\Gamma(e)]_{ij}^*Z_{\Gamma^{ij}}\ket{e}\\
			&=\frac{1}{|G|}\sum_{\Gamma}d^2_{\Gamma}\ket{e}\\
			&=\ket{e}
		\end{align}
		as expected.
		
		We thus have that prior to the application of any CMULT gates in \Eref{e:defcsgeneral} there are stabilizers for the system given by
		\begin{align}
			\tilde{S}^e(v)&=T_e(v)\mbox{, and}\\
			\tilde{S}_g^o(w)&=X_g^{\leftrightarrows}(w)
		\end{align}
		for every even site $v$ and odd site $w$, and all $g\in G$. In order to determine the stabilizers of the generalized cluster state then, we simply need to study the evolution of these operators under the action of controlled multiplication gates. This can be computed straightforwardly, and found to give
				
\begin{center}
\begin{tabular}{cc}
	$\mathrm{CMULT}^{\leftarrow}\qquad\qquad\qquad$&$\mathrm{CMULT}^{\rightarrow}$\\
      {$\!\begin{aligned} 
               X_g^{\leftarrow}\otimes I&\to X_g^{\leftarrow}\otimes X_g^{\leftarrow}\\
			X_g^{\rightarrow}\otimes I&\to \sum_hX_g^{\rightarrow}T_h\otimes X_{hg^{-1}h^{-1}}^{\leftarrow}\\
			I\otimes X_g^{\leftarrow}&\to \sum_hT_h\otimes X_{hgh^{-1}}^{\leftarrow} \\
			I\otimes X_g^{\rightarrow}&\to I\otimes X_g^{\rightarrow}\\
			T_g\otimes I&\to T_g\otimes I\\
			I\otimes T_g&\to \sum_hT_h \otimes  T_{hg}
			\end{aligned}$}
			\hspace{2cm}& \hspace{2cm}
	      {$\!\begin{aligned} 
			X_g^{\leftarrow}\otimes I&\to X_g^{\leftarrow}\otimes X_g^{\rightarrow}\\
			X_g^{\rightarrow}\otimes I&\to \sum_hX_g^{\rightarrow}T_h\otimes X_{hg^{-1}h^{-1}}^{\rightarrow}\\
			I\otimes X_g^{\leftarrow}&\to I\otimes X_g^{\leftarrow} \\
			I\otimes X_g^{\rightarrow}&\to \sum_hT_h\otimes X_{hgh^{-1}}^{\leftarrow}\\
			T_g\otimes I&\to T_g\otimes I\\
			I\otimes T_g&\to \sum_hT_h \otimes  T_{gh^{-1}}
			\end{aligned}$}
\end{tabular}
		\end{center}
		where the first tensor factor is the control qudit and the second the target.
		
	To construct the stabilizers of the generalized cluster states, we now build the state one CMULT gate at a time according to \Eref{e:defcsgeneral}. We will first develop the stabilizers for the even sites of the lattice, and then return and compute the stabilizers corresponding to odd sites.
		
		At each even site $v$ of the lattice, prior to the application of any CMULT gates the state is stabilized by $\tilde{S}^e(v)=T_e(i)$. This site will be the target of any CMULT gates applied to it. For each edge directed outwards from $v$, a $\mathrm{CMULT}^{\leftarrow}$ gate will be applied, and similarly a $\mathrm{CMULT}^{\rightarrow}$ gate will be applied for each inwards directed edge. In order to describe the evolution of $\tilde{S}^e(v)$, it will be convenient to define $n^{\leftarrow v}_w$ and $n^{\rightarrow v}_u$ as the neighbours of $v$ corresponding to the $w\th$ outward directed edge or $u\th$ inward directed edge respectively (where the ordering is given by $\#_v$).
		
	Recalling that $\mathrm{CMULT}^{\leftarrow}$ gates commute on common targets with $\mathrm{CMULT}^{\rightarrow}$ gates, we will consider first performing the $\mathrm{CMULT}^{\leftarrow}$ gates on site $v$. As we apply the CMULT gates on outward directed edges sequentially in the order specified by $\#_v$, the stabilizer for this site becomes
	\begin{align}
		\tilde{S}^e(v) &\to \sum_{g_1} T_{g_1}(v) T_{g_1}(n^{\leftarrow v}_1)  \\
		&\to  \sum_{g_1, g_2}T_{g_2g_1}(v) T_{g_1}(n^{\leftarrow v}_1)T_{g_2}(n^{\leftarrow v}_2) \\
		&\;\;\quad\vdots\nonumber\\
		&\to  \sum_{\substack{g_w \\ \forall w\leftarrow v}} T_{\prod_{w'} g_{w'}}(v)\prod_{w''}T_{g_{w''}}(n^{\leftarrow v}_{w''})
	\end{align}
	where in the final expression $w$ runs over all outward directed neighbours of $v$ and $\prod_{w} g_{w}= \cdots g_3g_2g_1$ (we also suppress the range of product and summation indices when they are clear from context). We can then similarly apply the gates on the inward directed edges of $v$ (again in the appropriate order) and obtain
	\begin{align}
		\tilde{S}^e(v) &\to  \sum_{\substack{h_u\\\forall u\rightarrow v}}\sum_{\substack{g_w\\ \forall w\leftarrow v}} T_{\left(\prod_{w'} g_{w'}\right)\left(\prod_{u'} h_{u'}\right)^{-1}}(v)\prod_{u''}T_{h_{u''}}(n^{\rightarrow v}_{u''})\prod_{w''}T_{g_{w''}}(n^{\leftarrow v}_{w''})\equiv S^e(v).\label{e:evenstaborig}
		\end{align}
			
	Since it now has support on sites other than $v$, one might now worry that we will also need to calculate how $\tilde{S}^e(v)$ transforms under the other CMULT gates acting on the neighbours of $v$. However, since these CMULT operations all commute with each other (they share a common control, not target), this is not a concern. Thus (\ref{e:oddstaborig}) is the stabilizer corresponding to each odd site of our cluster state, acting only on the given site and its nearest neighbours.

	We can also derive the stabilizers corresponding to odd sites of $\Lambda$ in a similar fashion. In this case, the stabilizers of an odd site $w$ prior to the application of any CMULT gates are given by
	\begin{align}
		\tilde{S}^o_g(w) &= X^{\leftrightarrows}_g(w)
	\end{align}
	for all $g$. Although both the left and right multiplication operators form valid stabilizers for an odd site, only one set is needed to specify the state completely. For simplicity we will choose to study the left multiplication stabilizers though the analysis for the right multiplication operators would proceed in an analogous way. With this in mind, we restrict to the case
	\begin{align}
		\tilde{S}^o_g(w) &= X^{\leftarrow}_g(w).
	\end{align}
	
	The site $w$ will act as the control for the CMULT gates applied to it. As such, it does not matter in which order the CMULT gates are applied at site $w$. However, in contrast to the even sites, the odd site stabilizers will be affected by the order of CMULT gates applied to the neighbouring (even) vertices.
	
	The CMULT gate corresponding to an outward directed edge from $w$ will act as $\mathrm{CMULT}^{\rightarrow}$ and inward directed edges correspond to $\mathrm{CMULT}^{\leftarrow}$. Applying a CMULT gate on an inward directed edge to neighbour $v$, we find $\tilde{S}^o_g(w)$ becomes
	\begin{align}
		\tilde{S}^o_g(w)\to X^{\leftarrow}_g(w)X^{\leftarrow}_g(v).
	\end{align}
	Now this operator has support on $v$ as well as $w$. This means that we must consider the effects of the CMULT gates acting not only between $w$ and its neighbours, but also between $v$ and its neighbours. In doing so, we will only need to consider the effects of those gates applied to $w$ after that corresponding to the edge $(w\leftarrow v)$. In addition, we need only consider edges with the same direction from $v$ (i.e.~outwards from $v$ in this case).
	
	We can make this effect explicit by finding the effect of these CMULT gates on an $X_g^{\leftarrow}$ operator acting on site $v$. As each successive $\mathrm{CMULT}^{\leftarrow}$ is applied to edges connecting $v$ to sites $\tilde{w}_1,\tilde{w}_2,\ldots$ that follow $w$ in the $\#_v$ order, this operator transforms as
	\begin{align}
		X_g^{\leftarrow}(v)&\to \sum_{h_1}X^{\leftarrow}_{h_1gh_1^{-1}}(v)T_{h_1}(\tilde{w}_1)\\
		&\to \sum_{h_1,h_2}X^{\leftarrow}_{(h_2h_1)g(h_2h_1)^{-1}}(v)T_{h_1}(\tilde{w}_1)T_{h_2}(\tilde{w}_2)\\
		&\;\;\quad\vdots\nonumber\\
		&\to \sum_{\substack{h_{\tilde{w}}\\\forall\tilde{w}\leftarrow v, \tilde{w}>w}}X^{\leftarrow}_{\left(\prod_{\tilde{w}'}h_{\tilde{w}'}\right)g\left(\prod_{\tilde{w}''}h_{\tilde{w}''}\right)^{-1}}(v)\prod_{\tilde{w}'''} T_{h_{\tilde{w}'''}}(\tilde{w}''')
	\end{align}
	where $\tilde{w}>w$ restricts to those $\tilde{w}$ that come after $w$ in $\#_v$.

This operator still acts as multiplication on site $v$, but by a group element conjugated by terms taking into account states of the neighbours of $v$. It will be convenient to define this operator for any site $v$ neighbouring $w$:
	\begin{align}
		\tilde{X}_g^{\leftarrow}(v|_w)\equiv \sum_{\substack{h_{\tilde{w}}\\\forall \tilde{w}\leftarrow v, \tilde{w}>w}}X^{\leftarrow}_{\left(\prod_{\tilde{w}'}h_{\tilde{w}'}\right)g\left(\prod_{\tilde{w}''}h_{\tilde{w}''}\right)^{-1}}(v)\prod_{\tilde{w}'''} T_{h_{\tilde{w}'''}}(\tilde{w}''').
	\end{align}

	Now having considered all the effects of site $v$ and its neighbours on the stabilizer for site $w$, we have the stabilizer
	\begin{align}
		\tilde{S}^o_g(w)\to X^{\leftarrow}_g(w)\tilde{X}^{\leftarrow}_g(v|_w).
	\end{align}
	
	As we apply this procedure to each of the edges directed inwards to $w$, the stabilizer transforms to
	\begin{align}
		\tilde{S}^o_g(w)\to X^{\leftarrow}_g(w)\prod_{v\rightarrow w}\tilde{X}^{\leftarrow}_g(v|_w).
	\end{align}
	
	In a similar way, we can also consider the effects of CMULT gates on outwards directed edges from $w$. If we apply a CMULT gate corresponding to an outward directed edge connecting site $w$ and $u$, the $X^{\leftarrow}(w)$ operator will be transformed by each gate as
	\begin{align}
		X^{\leftarrow}(w)\to X^{\leftarrow}(w)X^{\rightarrow}(u).
	\end{align}
	
	Following an analogous prescription to that used for the analysis of the inward directed edges, we are led to define the operator
	\begin{align}
		\tilde{X}_g^{\rightarrow}(u|_w)\equiv \sum_{\substack{h_{\tilde{w}}\\\forall\tilde{w}\rightarrow u, \tilde{w}>w}}X^{\rightarrow}_{\left(\prod_{\tilde{w}'}h_{\tilde{w}'}\right)g\left(\prod_{\tilde{w}''}h_{\tilde{w}''}\right)^{-1}}(u)\prod_{\tilde{w}'''} T_{h_{w'''}}(\tilde{w}''').
	\end{align}
	
	This allows us to compute the stabilizers corresponding to each odd site $w$ of a generalized cluster state as
	\begin{align}
		\tilde{S}^o_g(w)\to X^{\leftarrow}_g(w)\prod_{v\rightarrow w}\tilde{X}^{\leftarrow}_g(v|_w) \prod_{u\leftarrow w}\tilde{X}^{\rightarrow}_g(u|_w)\equiv S^o_g(w).\label{e:oddstaborig}
	\end{align}
	
	In contrast to the qubit cluster state, whose stabilizers only act on a site and its nearest neighbours, in general the operators (\ref{e:oddstaborig}) act on a site, its nearest neighbours, and also its next-nearest neighbours.
	
	Together, the even site stabilizers $S^e$ (\ref{e:evenstaborig}) and the odd site stabilizers $S^o$ (\ref{e:oddstaborig}) completely specify a generalized cluster state. The odd site stabilizers are monomial matrices in their current form, while the even site stabilizers are projectors and so can be made monomials by considering the alternative operators $2S^e-1$. This means that the generalized cluster states can be studied in the framework of monomial stabilizer groups~\cite{VandenNest2011} as claimed.
	
%------------------------------------------------------------------------------------------------------------%

\section{Properties of generalized cluster states}\label{s:properties}
	
	We will now explore some salient features of the generalized cluster states defined in the previous section. It is not our intention to provide a comprehensive specification of the properties of the states, but merely to comment on some features of the qubit cluster states and how they survive in the general case.
	
	One of the most important features of the qubit cluster state is that (on a suitable graph) it is a resource for universal measurement-based quantum computation~\cite{Raussendorf2003}. Since the qubit cluster state already has this universality property, considering generalized cluster states as resources for standard measurement-based quantum computation will not give any qualitative advantage in this sense. However, it is possible that generalized cluster states may yield advantages in terms of practical application, for example by finding more efficient implementations of non-Clifford gates or by having desirable spectral properties (when considered as ground states of Hamiltonians). We will not directly address these issues of practical advantage here, though some of the general properties we discuss may be relevant to any such study.
	
	\subsection{Measurements}\label{s:measure}
	
		The results of measurements in the standard qubit cluster states can easily be determined using the stabilizer formalism. The properties of the resulting states are largely insensitive to the measurement outcomes obtained. By comparison, measurements on generalized cluster states give rise to some qualitatively new phenomena that do not appear in the qubit cluster state, such as significant dependence of the output state on measurement outcomes.
		
\begin{figure}
\centering
\begin{subfigure}{0.4\textwidth}
\includegraphics{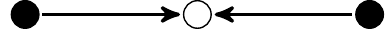}
%\begin{tikzpicture}[scale=3.5]
%	\def\off{0.05};
%	\def\rad{0.04};
%	\draw [->,>=stealth', line width=1] (0+\off,0)--(0.5-\off,0);
%	\draw [<-,>=stealth', line width=1] (0.5+\off,0)--(1-\off,0);
% 	\draw[black,fill=black] (1,0) circle (\rad);
% 	\draw[black,fill=black] (0,0) circle (\rad);
% 	\draw[black,fill=white] (0.5,0) circle (\rad);
%\end{tikzpicture}
\caption{}\label{f:lineclustera}
\end{subfigure}
\begin{subfigure}{0.4\textwidth}
\includegraphics{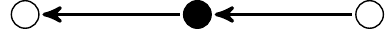}
%\begin{tikzpicture}[scale=3.5]
%	\def\off{0.05};
%	\def\rad{0.04};
%	\draw [<-,>=stealth', line width=1] (0+\off,0)--(0.5-\off,0);
%	\draw [<-,>=stealth', line width=1] (0.5+\off,0)--(1-\off,0);
% 	\draw[black,fill=white] (1,0) circle (\rad);
% 	\draw[black,fill=white] (0,0) circle (\rad);
% 	\draw[black,fill=black] (0.5,0) circle (\rad);
%\end{tikzpicture}
\caption{}\label{f:lineclusterb}
\end{subfigure}
\caption{Two different 3 qudit cluster states. Solid (open) circles represent even (odd) vertices. The simple structure of these graphs means that a vertex ordering is not required to specify the relevant cluster state. a) An even-odd-even 3 qudit cluster state. b) An odd-even-odd 3 qudit cluster state.}
\label{f:linecluster}
\end{figure}

		Ideally, we would like to understand the effect of measurements in the group element basis and the representation basis (the analogues of $Z$ and $X$ Pauli measurements) on an arbitrary generalized cluster state. However, since we can no longer use the simple Pauli stabilizer formalism to compute these effects, this general specification is outside the scope of this work. For this reason we will not exhaustively describe the results of measurement procedures on general cluster states. Instead, we will illustrate the kinds of new phenomena that can appear in a simple example. For this purpose, it is sufficient to consider 3 qudits on a line as in Fig.~\ref{f:lineclustera} or \ref{f:lineclusterb}.
	
	For a group $G$, these states are stabilized by the operators
	\begin{align}
	S(eoe,1)&=\sum_{g\in G}T_g\otimes T_g\otimes I,&
	S_g(oeo,1)&=X^{\leftarrow}_g\otimes X^{\leftarrow}_g\otimes I,&\\
	S_g(eoe,2)&=X^{\leftarrow}_g\otimes X^{\leftarrow}_g\otimes X^{\leftarrow}_g,&
	S(oeo,2)&=\sum_{g,h\in G}T_g\otimes T_{gh^{-1}}\otimes T_h,\\
	S(eoe,3)&=\sum_{g\in G}I\otimes T_g\otimes T_g\mbox{, and}&
	S_g(oeo,3)&=I\otimes X^{\rightarrow}_g\otimes X^{\rightarrow}_g,
	\end{align}
	where $eoe$ or $oeo$ represent the even-odd-even and odd-even-odd cluster states of Figs.~\ref{f:lineclustera} and \ref{f:lineclusterb} respectively.

	Explicitly then, these states can be written as analogues of the GHZ state in different bases:
		\begin{align}
		\ket{\mathcal{C}_{\mathrm{eoe}}}&=\frac{1}{\sqrt{|G|}}\sum_{g\in G}\ket{g}\otimes\ket{g}\otimes\ket{g}\\
		\ket{\mathcal{C}_{\mathrm{oeo}}}&=\frac{1}{|G|}\sum_{g,h\in G}\ket{g}\otimes\ket{gh^{-1}}\otimes\ket{h}.
	\end{align}
	
	The states $\ket{\mathcal{C}_{\mathrm{eoe}}}$ are clearly identical for any different choices of group $G$ with the same order. For a general graph of course this is not true (as with $\ket{\mathcal{C}_{\mathrm{oeo}}}$), but nonetheless we will see group structure arise in the analysis of measurement outcomes on $\ket{\mathcal{C}_{\mathrm{eoe}}}$. This is because the natural measurements to consider are analogues of Pauli $X$ and $Z$ operators in the qubit case, and these operators also inherit structure of the group $G$. 
	
	The two measurements we consider here are those in the group element basis $\{\ket{g}\}$ (corresponding to Pauli $Z$) and in the representation basis $\{\ket{\Gamma^{ij}}\}$ (corresponding to Pauli $X$). If we measure the central qubit of our cluster in these bases, we will find that some phenomena arise which have no counterpart in the qubit cluster state.
	
	Let us first recall what happens in the CSS qubit case (i.e.~$G=\Z_2$). Since the $eoe$ and $oeo$ qubit cluster states can be transformed into one another by Hadamard gates on each qubit, we need only consider one of these states (the behavior of the other can be determined by exchanging $X$ and $Z$). If we measure the central qubit of the $eoe$ cluster in the $Z$ basis and find outcome $m_z\in\Z_2$, the state on the remaining qubits becomes a product state $\ket{m_z}\otimes\ket{m_z}$. Alternatively, if we measure in the $X$ basis with measurement outcome $m_X\in\{\pm\}$, we find a maximally entangled state $\ket{0}\otimes\ket{0}+m_x\ket{1}\otimes\ket{1}$.
	
	For a general group $G$, the group element basis measurement proceeds in much the same fashion. For a outcome $m_g\in G$ of this measurement on an $eoe$ cluster state, the resulting state is given by the product state $\ket{m_g}\otimes\ket{m_g}$. Similarly, the $oeo$ cluster state is transformed to the maximally entangled state $\sum_{h\in G}\ket{h}\otimes\ket{h^{-1}m_g}$. In contrast, when performing a measurement in the representation basis, we find a qualitative departure from the qubit case. Such a measurement yields a triple of measurement outcomes $(m_{\Gamma},m_i,m_j)$ representing a matrix element of an irrep of $G$. Beginning with the $eoe$ cluster state, we note that it can be rewritten as
	\begin{align}
		\ket{\mathcal{C}_{\mathrm{eoe}}}&=\sum_{g}\sum_{\Gamma^{ij}}\frac{\sqrt{d_{\Gamma}}}{|G|}[\Gamma(g)]^*_{ij}\ket{g}\ket{\Gamma^{ij}}\ket{g}.
	\end{align}
	Thus if we measure the central qudit in the representation basis, the resulting state is given as
	\begin{align}
		\sum_{g}\sqrt{\frac{d_{m_\Gamma}}{|G|}}[m_\Gamma(g)]^*_{m_im_j}\ket{g}\ket{g}.\label{e:eoerepmeasure}
	\end{align}
	For any Abelian group, this will give a maximally entangled state as in the qubit case. However for a general group, the state (\ref{e:eoerepmeasure}) will be some less-than-maximally entangled state for any representation $m_{\Gamma}$ with dimension greater than $1$. We can see this by calculating the reduced density matrix of the first qudit as $\rho_{\mathcal{C}_{\mathrm{eoe}}}^{(1)}=\sum_{g}\frac{d_{m_\Gamma}}{|G|}|[m_\Gamma(g)]_{m_im_j}|^2\proj{g}$ which is not, in general, equal to $\sum_{g}\frac{1}{|G|}\proj{g}$. Of course, for any Abelian group all $d_{m_\Gamma}=1=|[m_\Gamma(g)]_{m_im_j}|^2$, and we recover the maximally entangled state as claimed.
	
	The analogous measurement on an $oeo$ cluster state produces a similar phenomenon. The resulting state after measurement is given by
	\begin{align}
		\ket{\mathcal{C}_{\mathrm{oeo}}}&\to \sum_{g,h}\frac{\sqrt{d_{m_\Gamma}}}{|G|}[m_\Gamma(gh^{-1})]^*_{m_im_j}\ket{g}\ket{h}\\
		&= \sum_{g,h\in G}\sum_{k=1}^{d_{m_\Gamma}}\frac{\sqrt{d_{m_\Gamma}}}{|G|}[m_\Gamma(g)]^*_{m_ik}[m_\Gamma(h^{-1})]^*_{km_j}\ket{g}\ket{h}\\
		&= \sum_{g,h\in G}\sum_{k=1}^{d_{m_\Gamma}}\frac{\sqrt{d_{m_\Gamma}}}{|G|}[\overline{m_\Gamma}(g)]_{km_i}[m_\Gamma(h)]_{m_jk}\ket{g}\ket{h}\\
		&= \sum_{k=1}^{d_{m_\Gamma}}\frac{1}{\sqrt{d_{m_{\Gamma}}}}\ket{\overline{m_\Gamma}^{km_i}}\ket{m_\Gamma^{m_jk}}
	\end{align}
for $\overline{m_{\Gamma}}$ the conjugate representation to $m_{\Gamma}$ (recall that we consider only unitary irreducible representations over $\C$). This leaves us with a maximally entangled state of dimension $d_{m_\Gamma}$. For Abelian groups with only 1-dimensional irreps, this gives the product state as in the qubit case. However, for general groups the behavior is non-trivial.

In particular, these phenomena mean that the property of maximal connectedness (i.e.~any two qudits in the state can be brought into a maximally entangled state with certainty by measurement) is not always present for a generalized cluster state as it is for the qubit cluster state.

In order to develop a standard measurement-based quantum computation protocol making use of the generalized cluster states, compensating for the subtle interplay between measurement outcomes and remaining entanglement would require tools outside the scope of standard cluster state computation methods. It may be possible to develop such a scheme explicitly by making use of techniques in Ref.~\cite{Gross2007}, combined with the PEPS representation of the generalized cluster states as found in~\Sref{s:peps}. However, we will not consider this question further in this work, instead focussing on the relationship between the generalized cluster states and the Kitaev quantum double models as discussed in Sections \ref{s:clusterqd} and \ref{s:newapps}. For this relationship, that the remaining entanglement in the state depends on prior measurement outcomes is crucial in reproducing the phenomenology of non-Abelian topological orders.

	\subsection{Global symmetries}\label{s:symmetry}
	
		The 1D cluster state is well known to have a global $\Z_2\times\Z_2$ symmetry that has a significant effect on its properties~\cite{Son2011,Son2012}. This allows it to be placed in the framework of symmetry-protected topological order (SPTO)~\cite{Gu2009,Chen2011,Schuch2011}. The symmetry group can also be shown to be related to the power and robustness of the cluster state as a measurement-based resource in both 1 and higher dimensions~\cite{Else2012, Else2012b}.
		
		The relevant global symmetry group of the infinite 1D CSS cluster state is represented by the operators
		\begin{align}
			U^o &= \prod_{s\;\mathrm{odd}}X(s)\mbox{, and}\\
			U^e &= \prod_{s\;\mathrm{even}}Z(s).
		\end{align}
		Since $U^o$ and $U^e$ clearly commute and are self-inverse, these operators generate a representation of $\Z_2\times\Z_2$.
		
\begin{figure}
\centering
\includegraphics{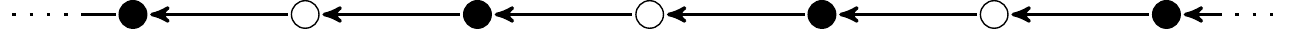}
%\begin{tikzpicture}[scale=3.5]
%	\def\off{0.05};
%	\def\rad{0.04};
%	\draw [-,>=stealth', line width=1] (-0.2+\off,0)--(0-\off,0);
%		\draw [-,loosely dotted, line width=1] (-0.4+\off,0)--(-0.2+\off,0);
%		%
%	\draw [<-,>=stealth', line width=1] (0+\off,0)--(0.5-\off,0);
%	\draw [<-,>=stealth', line width=1] (0.5+\off,0)--(1-\off,0);
%	\draw [<-,>=stealth', line width=1] (1+\off,0)--(1.5-\off,0);
%	\draw [<-,>=stealth', line width=1] (1.5+\off,0)--(2-\off,0);
%	\draw [<-,>=stealth', line width=1] (2+\off,0)--(2.5-\off,0);
%	\draw [<-,>=stealth', line width=1] (2.5+\off,0)--(3-\off,0);
%	\draw [<-,>=stealth', line width=1] (3+\off,0)--(3.2-\off,0);
%		\draw [-,loosely dotted, line width=1] (3.2-\off,0)--(3.4-\off,0);
%		%
% 	\draw[black,fill=black] (1,0) circle (\rad);
% 	\draw[black,fill=black] (0,0) circle (\rad);
% 	\draw[black,fill=black] (2,0) circle (\rad);
% 	\draw[black,fill=black] (3,0) circle (\rad);
% 	\draw[black,fill=white] (0.5,0) circle (\rad);
% 	\draw[black,fill=white] (1.5,0) circle (\rad);
% 	\draw[black,fill=white] (2.5,0) circle (\rad);
% 	%\draw[black,fill=white] (3.5,0) circle (\rad);
%\end{tikzpicture}
\caption{An infinite 1D generalized cluster state graph.}
\label{f:symmgraph}
\end{figure}

		Considering the analogous property of a generalized cluster state for arbitrary finite groups $G$, for simplicity we will restrict to the specific infinite 1D graph as shown in \Fref{f:symmgraph}.
		
	Explicitly, the cluster state corresponding to this graph can be written as
	\begin{align}
		\ket{\mathcal{C}_{\mathrm{line}}}&=\sum_{g_i}\cdots\ket{g_ig_{i+1}^{-1}}\ket{g_{i+1}}\ket{g_{i+1}g_{i+2}^{-1}}\ket{g_{i+2}}\ket{g_{i+2}g_{i+3}^{-1}}\ket{g_{i+3}}\ket{g_{i+3}g_{i+4}^{-1}}\cdots.\label{e:globsymclust}
	\end{align}

	It is clear by inspection that the global symmetries of this state can be written as
	\begin{align}
		U_g^o&= \prod_{s\;\mathrm{odd}}X^{\rightarrow}_g(s)\\
		U_{\Gamma}^e&= \frac{1}{d_{\Gamma}}\prod_{s\;\mathrm{even}}\sum_{i_s}Z_{\Gamma^{i_{s-2}i_s}}(s)
	\end{align}
	as can be directly verified. As in the qubit case the $U_g^o$ and $U_{\Gamma}^e$ commute trivially. The $U_g^o$ multiply as elements of $G$, and the $U_{\Gamma}^e$ transform as representations of $G$ (i.e.~$U_{\Gamma_1}^eU_{\Gamma_2}^e= U_{\Gamma_1\otimes\Gamma_2}^e$ and $U_{\Gamma_1}^e+U_{\Gamma_2}^e= U_{\Gamma_1\oplus\Gamma_2}^e$). Thus we deduce that the symmetry algebra of the generalized cluster state is given by the product of the group algebra and the dual (representation) algebra. For Abelian groups such as $\Z_2$, the representation algebra is isomorphic to that of the group itself, which recovers the familiar result for the qubit cluster state.
	
	Although the framework of symmetry protected topological order typically deals with states that have a group symmetry, we anticipate that many of the tools and results of SPTO may be extended to this more general setting.
	
	\subsection{PEPS representations}\label{s:peps}

		The qubit cluster state has an exact tensor network representation as a projected entangled pair state (PEPS)~\cite{Verstraete2004}. The PEPS ansatz~\cite{Affleck1988, Fannes1992, Hastings2006, Perez-Garcia2007a, Verstraete2006, Vidal2003a} is extremely useful for both analytical and numerical analysis of states. In particular, the description of the cluster state as a PEPS allows a reinterpretation of measurement-based quantum computation as a teleportation-based computation scheme~\cite{Verstraete2004} or in the correlation space picture~\cite{Gross2007}. Furthermore, it has also enabled an approximate 2-body parent Hamiltonian to be derived for an arbitrary cluster state~\cite{Bartlett2006, Griffin2008, Brellunpublishedpeps}. Here we show how the PEPS representation of a cluster state generalizes for an arbitrary finite group $G$.
	
	A PEPS is defined by an interaction graph $\Lambda$ and a set of ``projection'' maps $\mathcal{P}_s$ associated with each site of $\Lambda$. The state represented by the PEPS can be constructed by beginning with a set of $D$-dimensional maximally entangled ``virtual'' pairs $\sum_{i=1}^{D}\ket{i}_{mn}\ket{i}_{nm}\equiv\ket{\Phi_D}_{(m,n)}$ along each edge $(m,n)$ of $\Lambda$, with one qudit of each of these pairs associated with each of the vertices forming the edge (i.e.~the qudit labelled $mn$ is associated with vertex $m$ and vice versa). At each site $s\in\Lambda$, the projection map $\mathcal{P}_s$ is then applied to the qudits at each site, taking the combined $D^k$-dimensional virtual Hilbert space (for a site of valency $k$) to a $d$-dimensional ``physical'' Hilbert space. These physical qudits at each site of $\Lambda$ then typically correspond to the individual qudits of the state being represented. That is, the PEPS state is given by
	\begin{align}
	\ket{\psi_\mathrm{PEPS}}\equiv\bigotimes_{v\in \Lambda}\mathcal{P}(v)\bigotimes_{(m,n)\in \Lambda} \ket{\Phi_D}_{(m,n)}.
	\end{align}
	
	For the CSS qubit cluster state, the interaction graph $\Lambda$ is simply the (bipartite) graph specifying the cluster state. $D=d=2$ and the projection maps $\mathcal{P}$ are defined on odd and even sites of $\Lambda$ as~\cite{Verstraete2004}
	\begin{align}
		\mathcal{P}^o&=\ketbra{0}{0,\ldots ,0}+\ketbra{1}{1,\ldots ,1}\\
		\mathcal{P}^e&=\ketbra{+}{+,\ldots ,+}+\ketbra{-}{-,\ldots ,-}.
	\end{align}

	\newcommand{\sizedketbra}[2]{\left|{#1}\right\rangle\!\left\langle{#2}\right|}
	\newcommand{\sizedket}[1]{\left|{#1}\right\rangle}

For generalized cluster states, the PEPS interaction graph is again given by the cluster state graph. $D=d=|G|$ as might be expected, and the projection maps are given (up to normalization) by 
	\begin{align}
		\mathcal{P}^o&=\sum_{g}\ketbra{g}{g,\ldots ,g}\label{e:gpeps1}\\
		\mathcal{P}^e&=\sum_{g_1,\ldots,g_k,h_{k+1},\ldots,h_l}\sizedketbra{\left(\prod_{i}g_i\right)\left(\prod_{j}h_j\right)^{-1}}{g_1,\ldots ,g_k,h_{k+1},\ldots, h_{l}},\label{e:gpeps2}
	\end{align}
	where the $g_i$ represent the states of qudits corresponding to edges leaving the relevant even vertex, and the $h_j$ correspond to the edges entering the vertex. The products $\prod_{i}g_i$ and $\prod_{j}h_j$ are taken in the appropriate order specified by $\#$.

	In order to demonstrate that Eqns.~(\ref{e:gpeps1}-\ref{e:gpeps2}) define a PEPS for the generalized cluster states as claimed, it is useful to note that the effect of applying controlled multiplication gates on the physical level with target $\ket{e}$ can be reproduced by application of corresponding gates on the virtual level. We will distinguish typographically between $\mathrm{CMULT}$ gate acting on a physical target qudit and $\mathrm{cmult}$ gates acting on the virtual qudits.
	
	Consider $\mathrm{CMULT}$ gates acting on virtual control qudits and a common physical target qudit:
	\begin{align}
		\sideset{}{^\#}\prod_{i=1}^{l}\mathrm{CMULT}^{\leftrightarrows}(i,s)\ket{g_1,\ldots ,g_k,h_{k+1},\ldots, h_{l}}\ket{e}_s&=\ket{g_1,\ldots ,g_k,h_{k+1},\ldots, h_{l}}\sizedket{\left(\prod_{i}g_i\right)\left(\prod_{j}h_j\right)^{-1}}_s
	\end{align}
	where the $i$ index runs over the qudits grouped in the first ket. Compare this to $\mathrm{cmult}$ gates acting on virtual control and target qudits, before projecting the target qudits into a single physical qudit space:
	\begin{align}
		I \otimes \mathcal{P}^e_s\prod\mathrm{cmult}^{\leftarrow}&\ket{g_1,\ldots ,g_k,h_{k+1},\ldots, h_{l}}\otimes\ket{e,e,\ldots,e}_s= \nonumber\\
		&I\otimes\mathcal{P}^e_s\ket{g_1,\ldots ,g_k,h_{k+1},\ldots, h_{l}}\otimes\ket{g_1,\ldots ,g_k,h_{k+1},\ldots, h_{l}}_s\\
		&= \ket{g_1,\ldots ,g_k,h_{k+1},\ldots, h_{l}}\otimes\sizedket{\left(\prod_{i}g_i\right)\left(\prod_{j}h_j\right)^{-1}}_s\\
		&=\sideset{}{^\#}\prod_{i=1}^{l}\mathrm{CMULT}^{\leftrightarrows}(i,s)\ket{g_1,\ldots ,g_k,h_{k+1},\ldots, h_{l}}\ket{e}_s
	\end{align}
	where here only one $\mathrm{cmult}$ gate acts on each (virtual) control and each (virtual) target qudit. We can use this equivalence to give us the result that
	\begin{align}
		\ket{\psi_\mathrm{PEPS}}&=\left(\bigotimes_{w'\in \Lambda_o}\mathcal{P}^o(w')\bigotimes_{v'\in \Lambda_e}\mathcal{P}^e(v')\right)\bigotimes_{(w,v)\in \Lambda} \ket{\Phi_{|G|}}_{wv}\\
		&=\left(\bigotimes_{w'\in \Lambda_o}\mathcal{P}^o(w')\bigotimes_{v'\in \Lambda_e}\mathcal{P}^e(v')\right)\prod_{\substack{<m,n>\\ m\in \Lambda_o, n\in \Lambda_e}}\mathrm{cmult}^{\leftarrow}(m,n)
	 \left(\bigotimes_{w\in \Lambda_o}\ket{I,\ldots,I}_w\bigotimes_{v\in \Lambda_e}\ket{e,\ldots,e}_v\right)\\
	 &\propto \sideset{}{^\#}\prod_{\substack{<m,n>\\ m\in \Lambda_o, n\in \Lambda_e}}\mathrm{CMULT}^{\leftrightarrows}(m,n) \left(\bigotimes_{w\in \Lambda_o}\ket{I}_w\bigotimes_{v\in \Lambda_e}\ket{e}_v\right)\\
	 &= \ket{\mathcal{C}_{\Lambda,\#,G}}
	\end{align}
	as claimed, where on the penultimate line we used the facts that $\mathcal{P}^o_s$ commutes with CMULT on the control qudit, and $\ket{I}_s\propto \mathcal{P}^o_s \ket{I,\ldots, I}_s$.

	\subsection{Generalized cluster states and Kitaev quantum double states}\label{s:clusterqd}
	
		The qubit cluster state is related to the topologically ordered toric code in several ways. One important way is that the toric code can be defined by preparing a suitable cluster state and projecting (or measuring) a subset of the qubits into suitable states~\cite{Raussendorf2005}. Here we show that this relationship also extends between the generalized cluster states and the Kitaev quantum double models based on the same group. This relationship may also have practical application, for example in preparing certain quantum double states or in generalizing the topological cluster state computation protocol, as we will discuss in \Sref{s:newapps}.
	
\begin{figure}
\begin{center}
\begin{subfigure}{0.4\textwidth}
\includegraphics{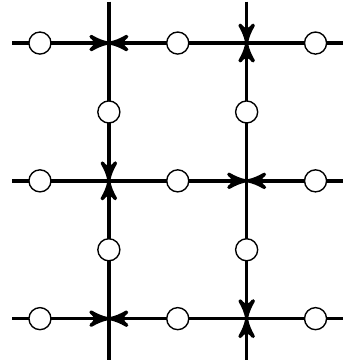}
%\begin{tikzpicture}[scale=1.4, baseline=1.25cm]
%	\def\off{0.3};	
%	
%	\draw[->,>=stealth', line width=1] (-0.5+\off,2)--(0.5,2);
%	\draw[<-,>=stealth', line width=1] (0.5,2)--(1.5,2);
%	\draw[-,>=stealth', line width=1] (1.5,2)--(2.5-\off,2);
%	\draw[-,>=stealth', line width=1] (-0.5+\off,1)--(0.5,1);
%	\draw[->,>=stealth', line width=1] (0.5,1)--(1.5,1);
%	\draw[<-,>=stealth', line width=1] (1.5,1)--(2.5-\off,1);
%	\draw[->,>=stealth', line width=1] (-0.5+\off,0)--(0.5,0);
%	\draw[<-,>=stealth', line width=1] (0.5,0)--(1.5,0);
%	\draw[-,>=stealth', line width=1] (1.5,0)--(2.5-\off,0);
%
%	\draw[<-,>=stealth', line width=1] (0.5,1)--(0.5,0);
%	\draw[<-,>=stealth', line width=1] (0.5,1)--(0.5,2);
%	\draw[->,>=stealth', line width=1] (1.5,1)--(1.5,0);
%	\draw[->,>=stealth', line width=1] (1.5,1)--(1.5,2);
%	
%	\draw[-,>=stealth', line width=1] (0.5,-\off)--(0.5,0);
%	\draw[-,>=stealth', line width=1] (0.5,2+\off)--(0.5,2);
%	\draw[->,>=stealth', line width=1] (1.5,-\off)--(1.5,0);
%	\draw[->,>=stealth', line width=1] (1.5,2+\off)--(1.5,2);
%	
%	\def\rad{0.08};	
%	\foreach \a in {0,2,4}{
% 		\foreach \b in {0,2,4}{
% 			\draw[black,fill=white] (\a*0.5,\b*0.5) circle (\rad);
%		}
%	}	
%	\foreach \a in {1,3}{
% 		\foreach \b in {1,3}{
% 			\draw[black,fill=white] (\a*0.5,\b*0.5) circle (\rad);
%		}
%	}	
%	
%\end{tikzpicture}
\caption{}\label{f:toricclusterdira}
\end{subfigure}
\qquad\qquad
\begin{subfigure}{0.4\textwidth}
\includegraphics{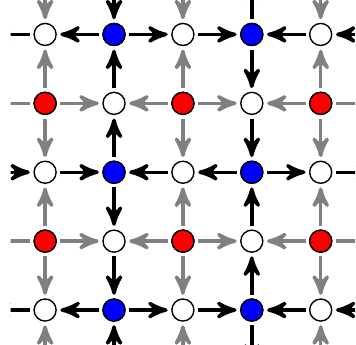}
\caption{}\label{f:toricclusterdirb}
\end{subfigure}
\end{center}
\caption{a) a choice of link directions for a quantum double model and b) a set of corresponding cluster state link directions. The qudits in the odd sublattice are shown as open circles, while the even sublattice is shown as colored circles. Qudits on the quantum double lattice reside on links. Red qudits in the cluster state are associated with plaquettes of the quantum double, while blue qudits are associated with stars.}
\label{f:toricclusterdirs}
\end{figure}

		Recall that a Kitaev quantum double model is defined on a directed lattice. For simplicity, we make a convenient canonical choice of edge direction (such that the edges around each plaquette either run clockwise or anticlockwise, as shown in \Fref{f:toricclusterdira}) and note that all alternative choices may be reached by local basis change. Quantum double states on these lattices have stabilizers~\cite{Kitaev2003}
		\begin{align}
			A(s)&=\sum_{g\in G}A_g(s)\mbox{ and}\label{e:qddef1}\\
			B(p)&=B_e(p)\label{e:qddeff}
		\end{align}
		for
		\begin{align}
			A_g^{h}(s)&=X^{\rightarrow}_g(s_U)X^{\leftarrow}_g(s_L)X^{\rightarrow}_g(s_D)X^{\leftarrow}_g(s_R)\\
			A_g^{v}(s)&=X^{\leftarrow}_g(s_U)X^{\rightarrow}_g(s_L)X^{\leftarrow}_g(s_D)X^{\rightarrow}_g(s_R)\\
			B_g^{CW}(p) &= \sum_{g_1g_2g_3g_4=g}T_{g_1}(p_U)T_{g_2}(p_L)T_{g_3}(p_D)T_{g_4}(p_R)\\
			B_g^{ACW}(p) &= \sum_{g_1^{-1}g_2^{-1}g_3^{-1}g_4^{-1}=g}T_{g_1}(p_U)T_{g_2}(p_L)T_{g_3}(p_D)T_{g_4}(p_R).
		\end{align}
		with $s_U,s_L,s_D,s_R$ and $p_U,p_L,p_D,p_R$ the neighbouring links located up, left, right, and down from the star $s$ or plaquette $p$ under consideration, where $CW$ or $ACW$ denotes whether the edges around the plaquette $p$ run clockwise or anticlockwise, and where $h$ or $v$ denote whether the horizontal or vertical edges point into the star $s$. The $A_g(s)$ and $B_e(p)$ commute pairwise, and the Kitaev quantum double ground state is defined as the common $+1$ eigenspace of all $A_g(s)$ and $B_e(p)$ operators. 
		
		In order to produce such a quantum double ground state, we begin with a generalized cluster state on a square lattice with half the lattice spacing of the quantum double lattice, as shown in \Fref{f:toricclusterdirb}. The even qudits of the generalized cluster state (shown as colored circles) will be projected into suitable states, while the odd sublattice qudits will form the space on which the quantum double state is defined. The red odd qudits correspond to plaquettes of the quantum double state, and as such can be labelled as either clockwise or anti-clockwise, depending on the direction of the links around the plaquette. Blue (odd) cluster state qudits correspond to vertices of the quantum double model, and can be labelled as horizontal or vertical depending on whether the horizontal or vertical incident links run into the vertex. An equivalent procedure can be found to project the odd qudits and retain the even qudits, but for simplicity we will treat only one case.
		
		The edge directions and vertex orderings of the cluster state lattice will have a significant effect on the final state. In particular, if we wish to recover a quantum double state on the lattice with edge directions as in \Fref{f:toricclusterdira}, one suitable choice of edge directions is shown in \Fref{f:toricclusterdirb}. Explicitly, the edges of the cluster state drawn in black in \Fref{f:toricclusterdirb} (i.e.~those that will form the quantum double lattice) should run in the opposite direction to the corresponding quantum double links. All grey edges run away from (odd) red sites. For each red site corresponding to an anticlockwise quantum double plaquette, the vertex ordering should be taken anticlockwise beginning from the topmost edge, while for each red site corresponding to a clockwise quantum double plaquette, the vertex ordering should be taken clockwise beginning from the topmost edge. The ordering of the edges around the blue sites is not particularly important, for simplicity we choose them anticlockwise beginning from the topmost edge. The origin of these convention choices may not appear particularly clear at this stage, but it will become apparent how they figure as we proceed in the analysis.
		
		The stabilizers for this generalized cluster state are given for red, blue, and odd sites respectively as
		\begin{align}
			S^{r,CW}(p) &= \sum_hT_h(p)\sum_{g_1g_2g_3g_4=h}T_{g_4}(p_U)T_{g_1}(p_L)T_{g_2}(p_D)T_{g_3}(p_R)\label{e:qdclustred1}\\
			S^{r,ACW}(p) &= \sum_hT_h(p)\sum_{g_3g_2g_1g_4=h}T_{g_4}(p_U)T_{g_1}(p_L)T_{g_2}(p_D)T_{g_3}(p_R)\label{e:qdclustred2}\\
			S^{b,v}(s) &= \sum_hT_h(s)\sum_{g_3g_1g_4^{-1}g_2^{-1}=h}T_{g_4}(s_U)T_{g_1}(s_L)T_{g_2}(s_D)T_{g_3}(s_R)\\
			S^{b,h}(s) &= \sum_hT_h(s)\sum_{g_2g_4g_1^{-1}g_3^{-1}=h}T_{g_4}(s_U)T_{g_1}(s_L)T_{g_2}(s_D)T_{g_3}(s_R)\\
			S^{o,u}(l) &= X^{\leftarrow}_g(l)\tilde{X}^{\leftarrow}_g(l_D|_l)\tilde{X}^{\leftarrow}_g(l_L|_l)\tilde{X}^{\rightarrow}_g(l_U|_l)\tilde{X}^{\rightarrow}_g(l_R|_l)\\
			S^{o,d}(l) &= X^{\leftarrow}_g(l)\tilde{X}^{\rightarrow}_g(l_D|_l)\tilde{X}^{\rightarrow}_g(l_L|_l)\tilde{X}^{\leftarrow}_g(l_U|_l)\tilde{X}^{\leftarrow}_g(l_R|_l)\\
			S^{o,l}(l) &= X^{\leftarrow}_g(l)\tilde{X}^{\leftarrow}_g(l_D|_l)\tilde{X}^{\leftarrow}_g(l_L|_l)\tilde{X}^{\rightarrow}_g(l_U|_l)\tilde{X}^{\rightarrow}_g(l_R|_l)\\
			S^{o,r}(l) &= X^{\leftarrow}_g(l)\tilde{X}^{\rightarrow}_g(l_D|_l)\tilde{X}^{\rightarrow}_g(l_L|_l)\tilde{X}^{\leftarrow}_g(l_U|_l)\tilde{X}^{\leftarrow}_g(l_R|_l)
		\end{align}
		where, as in the definition of the quantum double models, $CW$ and $ACW$ refer to the type of plaquette corresponding to the red site, $v$ and $h$ denote the type of star corresponding to the blue site, stabilizers associated to even sites have a $u$, $d$, $l$, $r$ designation according to the direction of the corresponding quantum double link, and vertices of the cluster state have neighbours denoted by $U$, $D$, $L$, and $R$ subscripts.
		
		To produce the Kitaev quantum double ground state from the generalized cluster state we have just described, we project the red even qudits (corresponding to plaquettes of the quantum double model) onto the $\ket{e}$ state, and the blue even qudits (corresponding to vertices) into the $\ket{I}$ state. In the case of the qubit cluster state, this corresponds to projections to the $\ket{0}$ or $\ket{+}$ states, respectively.
		
		Projecting even sites into $\ket{I}$ effectively removes them from the cluster state (indeed, we could alternatively have begun with a cluster state on a lattice lacking the blue sites). After performing just these blue site projections, the red site stabilizers are unchanged, the blue site stabilizers vanish and the odd site stabilizers are transformed to
		\begin{align}
			S^{o,u}_g(l) &\to X^{\leftarrow}_g(l)\tilde{X}^{\leftarrow}_g(l_D|_l)\tilde{X}^{\rightarrow}_g(l_U|_l)\label{e:qdclustnobluefirst}\\
			S^{o,d}_g(l) &\to X^{\leftarrow}_g(l)\tilde{X}^{\rightarrow}_g(l_D|_l)\tilde{X}^{\leftarrow}_g(l_U|_l)\\
			S^{o,l}_g(l) &\to X^{\leftarrow}_g(l)\tilde{X}^{\leftarrow}_g(l_L|_l)\tilde{X}^{\rightarrow}_g(l_R|_l)\\
			S^{o,r}_g(l) &\to X^{\leftarrow}_g(l)\tilde{X}^{\rightarrow}_g(l_L|_l)\tilde{X}^{\leftarrow}_g(l_R|_l)\label{e:qdclustnobluelast}
		\end{align}
		where the $\tilde{X}$ operators are now defined with respect to the lattice where the blue sites have been removed.
		
		If we then proceed with the projection of the red sites, the red plaquette stabilizers can easily be seen to evolve to
		\begin{align}
			S^{CW}(p) &\to \sum_{g_1g_2g_3g_4=e}T_{g_4}(p_U)T_{g_1}(p_L)T_{g_2}(p_D)T_{g_3}(p_R)\label{e:cwpstab}\\
			S^{ACW}(p) &\to \sum_{g_1^{-1}g_2^{-1}g_3^{-1}g_4^{-1}=e}T_{g_4}(p_U)T_{g_1}(p_L)T_{g_2}(p_D)T_{g_3}(p_R).\label{e:acwpstab}
		\end{align}
		
		It is less obvious, but products of the odd site stabilizers around each former blue site $s$ can be rearranged to form the following stabilizers after red site projection:
		\begin{align}
			S^{h}_g(s)&=X^{\rightarrow}_g(s_U)X^{\leftarrow}_g(s_L)X^{\rightarrow}_g(s_D)X^{\leftarrow}_g(s_R)\label{e:hstarstab}\\
			S^{v}_g(s)&=X^{\leftarrow}_g(s_U)X^{\rightarrow}_g(s_L)X^{\leftarrow}_g(s_D)X^{\rightarrow}_g(s_R).\label{e:vstarstab}.
		\end{align}
		
		This can most easily be seen by considering an explicit example of a cluster state in the neighbourhood of a single (removed) $h$-type blue site:
		\begin{align}
			\begin{tikzpicture}[scale=1.4, baseline=1.25cm]
				\def\off{0.11};
				\def\rad{0.08};
			 	\clip(-0.25,0.25) rectangle (1.25,1.75);
			 	 \draw [->,>=stealth', line width=1, color=gray] (0,0.5+\off)--(0,1-\off);
			  	\draw [<-,>=stealth', line width=1, color=gray] (0,1+\off)--(0,1.5-\off);
			  	 \draw [->,>=stealth', line width=1, color=gray] (\off,0.5)--(0.5-\off,0.5);
			  	\draw [<-,>=stealth', line width=1, color=gray] (0.5+\off,0.5)--(1-\off,0.5);
			 	\draw [->,>=stealth', line width=1, color=gray] (1,0.5+\off)--(1,1-\off);
			  	 \draw [<-,>=stealth', line width=1, color=gray] (1,1+\off)--(1,1.5-\off);
			  	 \draw [->,>=stealth', line width=1, color=gray] (1-\off,1.5)--(0.5+\off,1.5);
			  	\draw [<-,>=stealth', line width=1, color=gray] (0.5-\off,1.5)--(\off,1.5);
			  			\draw[black,fill=white] (0.5,0.5) circle (\rad);
			  			\draw[black,fill=white] (1,1) circle (\rad);
			  			\draw[black,fill=white] (0,1) circle (\rad);
			  			\draw[black,fill=white] (0.5,1.5) circle (\rad); 	
				\foreach \a in {0,2,4}{
			 		\foreach \b in {1,3}{
			 			\draw[black,fill=red] (\a*0.5,\b*0.5) circle (\rad);
			 			%\draw[black,fill=blue] (\b*0.5,\a*0.5) circle (\rad);
					}
				}
			\end{tikzpicture}
			\sim\;\;\sum_{g_1,g_2,g_3,g_4}
			\begin{tikzpicture}[scale=1.4, baseline=1.25cm]
				\def\off{0.11};
				\def\rad{0.08};
			 	\clip(-0.75,0.25) rectangle (1.75,1.75);
			  	\node at (0.5,0.5) {$\ket{g_2}$};
			  	\node at (1.2,1) {$\ket{g_3}$};
			  	\node at (-0.2,1) {$\ket{g_1}$};
			  	\node at (0.5,1.5) {$\ket{g_4}$}; 
			  	\node at (-0.2,0.5) {$\ket{g_2g_1}$};
			  	\node at (1.2,0.5) {$\ket{g_3g_2}$};
			  	\node at (-0.2,1.5) {$\ket{g_4g_1}$};
			  	\node at (1.2,1.5) {$\ket{g_3g_4}$}; 	
			\end{tikzpicture}.
		\end{align}
		This state is not stabilized by the appropriate operator (\ref{e:hstarstab}). However, if we then project the red sites into $\ket{e}$, we find:
			\begin{align}
			\sum_{g_1,g_2,g_3,g_4}\delta_{g_2g_1=e}\delta_{g_3g_2=e}\delta_{g_3g_4=e}\delta_{g_4g_1=e}
			\begin{tikzpicture}[scale=1.4, baseline=1.25cm]
				\def\off{0.11};
				\def\rad{0.08};
			 	\clip(-0.75,0.25) rectangle (1.75,1.75);
			  	\node at (0.5,0.5) {$\ket{g_2}$};
			  	\node at (1.2,1) {$\ket{g_3}$};
			  	\node at (-0.2,1) {$\ket{g_1}$};
			  	\node at (0.5,1.5) {$\ket{g_4}$}; 	
			\end{tikzpicture}
		\end{align}	
		which is indeed stabilized by $S^{h}_g$ as claimed. Note that if we had chosen a projection such that, for example, $g_3g_2=h$ for some $h\neq e$, then this would no longer be true. If instead, we projected into $\sum_{h\in C}\ket{h}$ for some conjugacy class $C$, $S^h_g$ would again stabilize the state. In the language of the quantum double anyons, this is due to the fact that only uniform superpositions over conjugacy classes correspond to pure magnetic type excitations.
	
		The stabilizers we just derived (\ref{e:cwpstab}-\ref{e:vstarstab}) are precisely the $A_g(s)$ and $B_e(p)$ stabilizers of the quantum double (\ref{e:qddef1}-\ref{e:qddeff}). Thus we have a procedure to define the topologically ordered Kitaev quantum double states by a sequence of projections on the generalized cluster states. 
		
			One might hope that that for any scheme which builds some interesting state by measurement of $Z$ and $X$ on a (bipartite) qubit cluster state, we could define a generalized model for a finite group $G$ built by performing suitable projections on the generalized cluster state as we have done here. These models may or may not be able to be actually constructed from a cluster state in general by measurement (their interesting properties may or may not survive the complications of non-ideal measurement outcomes, as will be mentioned in \Sref{s:practical}), but simply being able to define such models may be of independent interest. An example of a model that can be defined in such a way is a non-Abelian generalization of the color codes, presented in Ref.~\cite{Brellunpublishedcolor}. One should also note the generality (and hence complexity) of using the projection procedure in this way. We are able to define many generalizations within this same framework by choosing different even and odd sublattices, different link directions, and different link orderings at vertices. Furthermore, alternative projections can be chosen to generalize the measurement of the $Z$ or $X$ operators, as is done in Ref.~\cite{Brellunpublishedcolor}. In general, the properties of the resulting state will be significantly affected by the choices of these parameters.
			
%------------------------------------------------------------------------------------------------------------%			

\section{Discussion}\label{s:discuss}

	\subsection{Applications}\label{s:newapps}
		
		\subsubsection{Producing the quantum double states}\label{s:practical}
		
		In practice, one may be interested in using the relationship between the generalized cluster states and quantum double models discussed in \Sref{s:clusterqd} as a resource to prepare quantum double states in the laboratory, by replacing the projections by suitable measurements. In the case of the toric code, replacing $\ket{0}$ and $\ket{+}$ projections with measurements in the $Z$ and $X$ bases respectively does not significantly affect the properties of the resulting state. However, in general the effects outlined in \Sref{s:measure} mean that the situation is not so simple.
		
		To illustrate some of the resulting phenomena, consider replacing the $\ket{e}$ projections at each red site in \Fref{f:toricclusterdirb} by measurements in the group element basis. After projection of the blue sites of the cluster state shown in \Fref{f:toricclusterdirb} into the $\ket{I}$ state (equivalently, removal of the blue sites), the stabilizers of the state corresponding to the red sites are given as in equations (\ref{e:qdclustred1}-\ref{e:qdclustred2}), while the remaining odd site stabilizers are as in (\ref{e:qdclustnobluefirst}-\ref{e:qdclustnobluelast}). The effect on the red site stabilizers of measuring these sites in the group element basis is clear: with measurement outcomes $\{m_p\}$, the stabilizers transform to
		\begin{align}
			S^{CW}(p) &= \sum_{g_1g_2g_3g_4=m_p}T_{g_4}(p_U)T_{g_1}(p_L)T_{g_2}(p_D)T_{g_3}(p_R)\label{e:qdmeasureclustfirst}\\
			S^{ACW}(p) &= \sum_{g_3g_2g_1g_4=m_p}T_{g_4}(p_U)T_{g_1}(p_L)T_{g_2}(p_D)T_{g_3}(p_R).\label{e:qdmeasureclustlast}
		\end{align}
		
		The effect on the odd site stabilizers is less obvious. As in the previous section where red sites were projected into $\ket{e}$, there will be a stabilizer corresponding to each star of the lattice. These can also be straightforwardly calculated by considering a single isolated vertex and its four neighbouring plaquettes.	
		
		We can interpret the state after measurement of the blue sites as a superposition of excited states of quantum double anyons. Explicitly, anyons in the quantum double models can be created and transported by so-called ribbon operators $F^{(h,g)}_{\rho}$ corresponding to pairs of group elements $(h,g)$ and fattened paths (ribbons) $\rho$ on the lattice~\cite{Kitaev2003,Bombin2008b}. The ends of ribbons, ``sites'', are composed of a neighbouring vertex-star pair. Ribbon operators create anyonic charges at each end of the ribbon $\rho$. In particular, the operator $F^{(m_p,e)}_{\rho}$ ending on an appropriate site will transform a stabilizer of the form (\ref{e:cwpstab}) or (\ref{e:acwpstab}) into (\ref{e:qdmeasureclustfirst}) or (\ref{e:qdmeasureclustlast}) respectively. Similarly, the star-type stabilizers can be obtained by transforming the naive stabilizers (\ref{e:hstarstab}) and (\ref{e:vstarstab}) under $F^{(m_p,e)}_{\rho}$ on suitable ribbons. Further, if we replaced all blue site projections with measurements in the representation basis, we could interpret the remaining state in a similar way.
		
		In some senses, this is a trivial statement, as all states on the lattice (up to boundary conditions) are superpositions of some excited states of the quantum double model. However, using this interpretation may be fruitful for example in a detailed analysis of the topological cluster state computation protocol sketched in the following section.
				
			\subsubsection{Adiabatic topological cluster state quantum computation}\label{s:atcsqc}

	Apart from standard measurement-based quantum computation, the qubit cluster state can also be used to implement topological cluster state quantum computation~\cite{Raussendorf2007} (TCSQC). This is a computation scheme that combines practical advantages of measurement-based quantum computation with some natural robustness to noise from topological computation schemes. It is based on the relationship between the cluster states and the toric code discussed and generalized in \Sref{s:clusterqd}.

	In the TCSQC protocol, measurements are performed on a cubic lattice cluster state. By interpreting one direction of the lattice as a ``simulated time'' direction, the resulting procedure can be interpreted as the evolution of punctures in the surface of the toric code in simulated time. The world-lines of the punctures are determined by the chosen measurement settings, while the measurement outcomes may be interpreted as specifying world-lines of toric code anyons, which can then be compensated for by classical post-processing. Though this scheme enjoys the native robustness that accompanies topological computation, the gates that can be performed in this way are insufficient for universal quantum computation, and so must be supplemented with non-topological operations whose fault-tolerance is guaranteed separately.
	
	It is an interesting question as to whether a similar measurement-based computation scheme exists that enjoys universal topological quantum computation. An obvious candidate strategy is to attempt to replace the simulated toric code punctures with simulated defects in another more complicated topological model, such as the quantum doubles. If we were to construct an analogous procedure using the generalized cluster states for a non-Abelian group $G$, appropriately generalizing the qubit TCSQC protocol would not generally succeed for the reason that the effect of random measurement outcomes could not be efficiently classically processed. Random measurement outcomes would be interpreted as (possibly superpositions of) anyons, and their dynamics - and hence effect on the logical state - is unlikely to be efficiently classically calculable in general (particularly if the anyon dynamics is BQP complete).
	
	One possible way to salvage such a generalized topological cluster state computation scheme is by removing the measurements from the protocol altogether. Adiabatic cluster state computation~\cite{Bacon2010,Bacon2013,Antonio2013} makes use of the cluster state as a modular prototype for computation by adiabatic deformation of a Hamiltonian. A standard measurement-based cluster state computation may be translated to the adiabatic setting by simulating the measurement with the adiabatic application of a strong field in the direction of the desired measurement result. Thus in this scheme, the analogue of measurement results are not random, and so need not be compensated for by classical post-processing.
	
	Though the adiabatic cluster state computation approach is typically applied to a standard cluster state computation, we could equally well apply it to the topological cluster state scheme. In this setting, the choice of fields would effectively set the world-lines of punctures and anyons in simulated time, without randomness. Making use of these techniques, it should be possible to perform universal topological cluster state computation with the generalized cluster states by simulating the evolution of punctures (as with the standard TCSQC scheme) in a sufficiently complicated Kitaev quantum double model\footnote{Though computation by braiding punctures in non-Abelian anyon systems has not been extensively studied, the fact that punctures contain anyonic charges and the braiding of anyonic charges can be sufficient for universal quantum computing~\cite{Kitaev2003,Mochon2003,Mochon2004} suggests that the braiding of punctures should also be sufficient for universal quantum computation in general.}. Alternatively, since the world-lines of anyons may now be controlled directly by choosing the adiabatic fields appropriately, and since braiding of suitable anyons is sufficient to implement universal quantum computation~\cite{Kitaev2003,Mochon2003,Mochon2004}, it should also be possible to compute in a generalized topological cluster state computation scheme by adiabatically simulating the evolution of a suitable class of anyons. These methods for adiabatic topological cluster state computation may be contrasted with the direct approach of adiabatic topological quantum computation~\cite{Cesare2014}.

	\subsubsection{Universal topological measurement-based quantum computation}
	In the scheme laid out in \Sref{s:atcsqc} for adiabatic topological quantum computation with these generalized cluster states, it was in part because of the infeasibility of classical processing that standard measurement-based computing techniques could not be used. The universality of quasiparticle braiding in the relevant anyon model meant that it was infeasible to calculate the effect of any stray braids caused by undesirable measurement outcomes (the most naive implementation of the scheme would in fact produce superpositions of anyon states which would generally be non-simulable in any case, but we believe that it may be possible to circumvent this with a more sophisticated protocol in some cases).
	
	However, recall that in the qubit topological cluster state computation scheme the computation proceeds not by simulating anyon braiding, but by simulating the braiding of punctures in the surface of the system. It is not clear that the computational power of such puncture braiding and the computational power of anyon braiding in such a model must always coincide. In fact, for another type of topological defect known as a twist, braiding of these objects can be universal for quantum computation while the braiding of the relevant anyons remains classically simulatable~\cite{Barkeshli2013}. Thus it may be possible to find some anyon model where braiding of punctures is universal, while braiding of anyons is classically simulatable. Although the interaction between topological defects and anyons can be quite complex, this could conceivably lead to a universal topological measurement-based quantum computation scheme based on an appropriate generalized cluster state (whether as described in this paper or from the possible extensions noted below).

	\subsection{Extensions}\label{s:extension}
	
		Although as presented here, the generalized cluster states are based on finite group algebras, there is little in the construction that is particularly restricted to finite groups. Several related or further generalizations of this construction suggest themselves.
		
		The clearest example is the extension from finite groups to Lie groups. In particular, by considering the group of real numbers under addition, our construction reproduces the continuous-variable cluster states~\cite{Zhang2006, Menicucci2006}. This gives a unified framework for all previously known variants of the cluster state. This will be presented elsewhere~\cite{Brellunpublishedlie}.
		
		As this construction was inspired by the generalization of the toric code to the quantum double models based on finite groups, it is natural to ask if the later generalizations of the toric code to arbitrary finite-dimensional Hopf $C^*$ algebras~\cite{Buerschaper2013} and the conjectured generalization to weak Hopf algebras~\cite{Buerschaper2013,Buerschaper2013b} could also be applied to generalized cluster states. There is an obstacle in directly extending the construction to Hopf algebras, as the natural generalization of the controlled multiplication operation would be according to the coproduct $\Delta$ of the Hopf algebra. This is the operation that maps from one system to a tensor product of two systems.
		
		The coproduct of a group algebra is simply $\Delta(g)=g\otimes g$ for $g\in G$. In a general Hopf algebra, the coproduct of an element is given by $\Delta(a)=\sum_i a_{(1)}^{(i)}\otimes a_{(2)}^{(i)}$. The natural generalization of the controlled multiplication gate is defined using this structure, e.g.~$\mathrm{CMULT}\ket{a}\ket{b}=\sum_i \ket{a_{(1)}^{(i)}} \ket{a_{(2)}^{(i)}b}$. However, this gate no longer need commute on control qudits. For this reason, if we were to use this CMULT gate to construct a cluster state, the circuit would no longer be finite-depth and the stabilizer operators derived in analogy to those in \Sref{s:cs} would no longer be finite weight in general.
		
		One way to define cluster states based on Hopf algebras that can be constructed by constant depth circuits and have local stabilizers would be to restrict the graphs on which these states are defined. In particular, consider a graph $\Lambda$ that is bipartite and every site $v$ in the even sublattice $\Lambda_e$ is 2-valent. Then one edge incident to $v$ could act as controlled left multiplication and the other edge acts as controlled right mutliplication. Since these two operations always commute on common targets, this construction would yield a state with the desired properties. Of course choosing graphs of this kind is a very restrictive constraint, and so it would be interesting to see if an alternative mechanism to retain locality is possible.
		
		Apart from Lie groups and Hopf algebras as motivated by the Kitaev quantum double models, it would also be interesting to study generalizations of the cluster states motivated by the Levin-Wen string net models~\cite{Levin2005}. Instead of a group, these models are specified by a fusion category. The simplest choice gives the toric code as in the quantum double models. As in Table \ref{t:css-qd}, we could build qudits and associated operator algebras inheriting features of a given fusion category. It would be interesting to develop cluster states for a given fusion category in this way.
	
	Finally, we note that in some of these extensions it may be possible to construct some cluster states that are related to the standard definition (\ref{e:qubitcirc}) instead of the CSS definition (\ref{e:csscirc}). This should be possible when considering algebraic objects which are self-dual in the relevant sense (as with Abelian groups).
	
	\subsection{Broader implications}	
	
		As noted in \Sref{s:symmetry}, the generalized cluster states on an infinite chain extend the standard notion of symmetry protected states with a symmetry group to states that simply have a symmetry algebra. The consequences of this are not clear, and it may be of interest to consider the notion of symmetry protected phases that are labelled by more general objects than groups.
	
		The Pauli stabilizer formalism has proved spectacularly successful at describing a wide variety of states, and making them amenable to both analytical and numerical study. Several similar constructions have been proposed, including the monomial stabilizer formalism~\cite{VandenNest2011} mentioned earlier, among others~\cite{Gottesman1999, Ni2014}. It is known that all Pauli stabilizer states are equivalent (under local Clifford circuits) to a qubit cluster state~\cite{VandenNest2004}. It may be of interest to determine whether a similar family of states can be defined by local equivalence to a generalized cluster state for any given group $G$. Although these all fall under the umbrella of monomial stabilizer states, it may be advantageous to define smaller classes that have more structure than the general case. 

%------------------------------------------------------------------------------------------------------------%

\begin{acknowledgments}
	We thank Stephen Bartlett, Andrew Doherty, Dominic Else, Steve Flammia, and Pieter Naaijkens for fruitful discussions and helpful comments. This work is supported by the ARC via the Centre of Excellence in Engineered Quantum Systems (EQuS) project number CE110001013, by the ERC grant QFTCMPS, and by the cluster of excellence EXC 201 Quantum Engineering and Space-Time Research.
\end{acknowledgments}

%------------------------------------------------------------------------------------------------------------%

\end{document}